\documentclass[prd,superscriptaddress,nofootinbib,amsmath,amssymb,aps,11pt]{revtex4}

\usepackage{bm}
\usepackage{amsfonts}
\usepackage{latexsym}
\usepackage[latin1]{inputenc}
\usepackage{graphicx}
\usepackage{amsmath}
\usepackage{palatino}
\usepackage{mathpazo}
\usepackage[normalem]{ulem}

\usepackage{booktabs}
\usepackage{dcolumn}

\def\jnl@style{\it}
\def\aaref@jnl#1{{\jnl@style#1}}

\def\aaref@jnl#1{{\jnl@style#1}}

\def\aj{\aaref@jnl{AJ}}                   
\def\apj{\aaref@jnl{ApJ}}                 
\def\apjl{\aaref@jnl{ApJ}}                
\def\apjs{\aaref@jnl{ApJS}}               
\def\apss{\aaref@jnl{Ap\&SS}}             
\def\aap{\aaref@jnl{A\&A}}                
\def\aapr{\aaref@jnl{A\&A~Rev.}}          
\def\aaps{\aaref@jnl{A\&AS}}              
\def\mnras{\aaref@jnl{Mon.~Not.~Roy.~Astron.~Soc.}}             
\def\prd{\aaref@jnl{Phys.~Rev.~D}}        
\def\prc{\aaref@jnl{Phys.~Rev.~C}}  
\def\prl{\aaref@jnl{Phys.~Rev.~Lett.}}    
\def\qjras{\aaref@jnl{QJRAS}}             
\def\skytel{\aaref@jnl{S\&T}}             
\def\ssr{\aaref@jnl{Space~Sci.~Rev.}}     
\def\zap{\aaref@jnl{ZAp}}                 
\def\nat{\aaref@jnl{Nature}}              
\def\aplett{\aaref@jnl{Astrophys.~Lett.}} 
\def\apspr{\aaref@jnl{Astrophys.~Space~Phys.~Res.}} 
\def\physrep{\aaref@jnl{Phys.~Rep.}}      
\def\physscr{\aaref@jnl{Phys.~Scr}}       
\def\commat{\aaref@jnl{Comm.~Math.~Phys.}}              
\def\science{\aaref@jnl{Science}}               
\def\cqg{\aaref@jnl{Classical Quant.~Grav.}}            
\def\jpcs{\aaref@jnl{JPCS}}                                     
\def\ijmpd{\aaref@jnl{Int.~J.~Mod.~Phys.~D}}                    
\def\grg{\aaref@jnl{Gen.~Relat.~Gravit.}}               
\def\rpp{\aaref@jnl{Rep.~Prog.~Phys.}}          
\def\npa{\aaref@jnl{Nucl.~Phys.~A}}        
\def\lrr{\aaref@jnl{Living Rev.~Rel.}}                   
\def\jcap{\aaref@jnl{J.~Cosmology Astropart.~Phys.}}    
\def\rmp{\aaref@jnl{Rev.~Mod.~Phys.}}   

\allowdisplaybreaks[1]

\begin{document}

	\title{Multi-scalar Gauss-Bonnet gravity -- hairy black holes and scalarization }
	
	\author{Daniela D. Doneva}
	\email{daniela.doneva@uni-tuebingen.de}
	\affiliation{Theoretical Astrophysics, Eberhard Karls University of T\"ubingen, T\"ubingen 72076, Germany}
	\affiliation{INRNE - Bulgarian Academy of Sciences, 1784  Sofia, Bulgaria}
	
	\author{Kalin V. Staykov}
	\email{kstaykov@phys.uni-sofia.bg}
	\affiliation{Department of Theoretical Physics, Faculty of Physics, Sofia University, Sofia 1164, Bulgaria}
	
	\author{Stoytcho S. Yazadjiev}
	\email{yazad@phys.uni-sofia.bg}
	\affiliation{Theoretical Astrophysics, Eberhard Karls University of T\"ubingen, T\"ubingen 72076, Germany}
	\affiliation{Department of Theoretical Physics, Faculty of Physics, Sofia University, Sofia 1164, Bulgaria}
	\affiliation{Institute of Mathematics and Informatics, 	Bulgarian Academy of Sciences, 	Acad. G. Bonchev St. 8, Sofia 1113, Bulgaria}
	
	\author{Radostina Z. Zheleva}
	\email{radostinazheleva94@gmail.com}
	\affiliation{Department of Theoretical Physics, Faculty of Physics, Sofia University, Sofia 1164, Bulgaria}


	\begin{abstract}
		In the present paper we consider multi-scalar extension of Einstein-Gauss-Bonnet gravity. We focus on multi-scalar  Einstein-Gauss-Bonnet models  whose target space is a three-dimensional maximally symmetric  space, namely either $\mathbb{S}^3$, $\mathbb{H}^3$ or $\mathbb{R}^3$, and in the case when the map  $\text{\it spacetime} \to \text{\it target space}$ is nontrivial. We prove numerically the existence of black holes in this class of models for several Gauss-Bonnet coupling functions, including the case of scalarization. We also  perform systematic study of a variety of black hole characteristics and  the space-time around them, such as the area of the horizon, the entropy and the radius of the photon sphere. One of the most important properties of the obtained solutions is that the scalar charge is zero and thus the scalar dipole radiation is suppressed which leads to much weaker observational constraints compared to the majority of modified theories possessing a scalar degree of freedom. For one of the coupling functions we could find branches of scalarized black holes which have a nontrivial structure -- there is non-uniqueness of the scalarized solutions belonging to a single branch and there is a region of the parameter space where most probably stable scalarized black holes coexist with the stable Schwarzschild black holes. Such a phenomena can have a clear observational signature.
	\end{abstract}

	\maketitle
	
	\section{Introduction}
	The unifying theories   predict one or more scalar partners of the tensor graviton. The scalar degrees of freedom 
	are usually coupled to the curvature invariants of spacetime \cite{Barack_2019, Berti_2015}. A notable example is the Einstein-scalar-Gauss-Bonnet (ESGB) gravity. In this theory  the scalar degree is coupled to the Gauss-Bonnet invariant and the field equations are of second differential order as in general relativity (GR). ESGB gravity with only one dynamical scalar field  and different coupling functions has recently  attracted a lot of interest.
	A particular class of ESGB theories is the Einstein-dilaton-Gauss-Bonnet gravity whose coupling function is exponential. Various aspects of black holes in this model were studied in a number of papers \cite{Mignemi_1993}--\cite{Bakopoulos_2020} including their quasinormal modes \cite{Blazquez-Salcedo:2016enn,Blazquez-Salcedo:2017txk}. The ESGB gravity with more general coupling functions were studied in   \cite{Antoniou_2018}--\cite{Bakopoulos_2019a}.  
	
	It was recently shown  in \cite{Doneva_2018,Silva_2018} that in a certain class of ESGB theories and in the extreme curvature regime there exist new black hole solutions which are formed by spontaneous scalarization of the Schwarzschild black holes. In this regime the Schwarzschild solution becomes unstable below certain mass, and new branches of solutions with nontrivial scalar field bifurcate from the Schwarzschild one. This scalarization is induced by the curvature of the spacetime in contrast with the spontaneous scalarization of neutron stars \cite{Damour_1993} and black holes \cite{Stefanov_2008, Doneva_2010} in the scalar-tensor theories, which is induced by the presence of matter.  It was further shown that for certain ranges of the parameters and choices of the coupling function, these solutions are stable  \cite{Salcedo_2018}--\cite{Macedo:2019sem}. The extension to rapid rotation was done in \cite{Kunha_2019,Collodel2019} and other scalarized black holes in Gauss-Bonnet gravity were considered in \cite{Doneva_2018a}--\cite{Hod:2019pmb}. The spontaneous scalarization and the scalarized black holes in a Horndeski type generalization of ESTGB gravity was studied in \cite{Andreou:2019ikc}--\cite{Ventagli_2020} while other theories were addressed in \cite{Brihaye_2019}--\cite{Ikeda:2019okp}.

	In the present paper we consider a multi-scalar extension of ESGB gravity. As in the multi-scalar-tensor
	theories \cite{Damour_1992,Horbatsch_2015}, instead of a single scalar field we introduce $N$ dynamical scalar fields $\varphi^a$ coupled to the Gauss-Bonnet invariant 
	and taking values in an abstract Riemannian target space. The presence of many scalar fields is not just quantitative increase 
	of the scalar degrees of freedom -- it can change the picture drastically.  It was recently  shown within the framework of the multi-scalar-tensor
	theories, that new types of compact objects can exist due to the presence of multiple scalar degrees of freedom when the map 
	$\varphi : \text{\it spacetime} \to \text{\it target space}$ generated by the scalar fields is nontrivial \cite{Yazadjiev_2019}--\cite{Doneva:2020afj}. The first steps towards exploring the linear stability of such objects were made in  \cite{Doneva:2020csi}, there it was shown that a completely new class of neutron stars possessing nonzero topological charge, is stable against linear perturbations.

	In  this work we prove numerically the existence of static and spherically symmetric  black holes in certain classes of  multi-scalar-Einstein-Gauss-Bonnet (MSEGB) gravity for linear and exponential coupling functions, as well as coupling functions leading to scalarizaions. More precisely we  consider MSEGB gravity whose target  space is a 3-dimensional maximally symmetric space, namely $\mathbb{S}^3$, $\mathbb{H}^3$ or $\mathbb{R}^3$ and in the case of a nontrivial map $\varphi : \text{\it spacetime} \to \text{\it target space}$. We also perform systematic study of many black hole characteristics  such as the area of the horizon, the entropy, and the photon sphere. 
	
	We should note that black holes with nontrivial scalar field, especially in the case of scalarization, might lead to tension with cosmology that can be cured in certain cases but not always \cite{Alby:2017dzl}--\cite{Silva:2019rle}. The problem with the cosmological instability, though, requires more profound investigation and moreover it was never addressed in more complicated cases such as the multi-scalar theories which offer much richer phenomenology and possibilities to circumvent different problems. Nevertheless, in the present paper we consider the MSEGB  gravity as an effective model operating only on astrophysical scales without requiring it to be a complete theory explaining the accelerated expansion phenomena or the early Universe. 
	
	In Section II we give briefly the mathematical formulation of multi-scalar-Einstein-Gauss-Bonnet gravity, derive the relevant reduced field equations and discuss the criteria for the existence of black holes. The obtained numerical solutions are presented in Sec. III divided into two major subsections -- one for hairy black holes with linear and exponential coupling, and one for scalarized solutions with two different forms of the coupling function. The paper ends with Conclusions where some observational perspectives are also discussed.

	\section{Multi-scalar Gauss-Bonnet gravity and black holes}
	
	Let us begin with  a precise description of the  MSEGB gravity. We consider a 4-dimensional spacetime ${\cal M}$ supplemented with spacetime metric $g_{\mu\nu}$  and additional  $N$ scalar fields $\varphi^{a}$  which take value in a coordinate patch of an N-dimensional Riemannian (target) manifold ${\cal E}_{N}$  with (positively definite) metric $\gamma_{ab}(\varphi)$ defined on it \cite{Damour_1992,Horbatsch_2015}.  From a more global point of view $\varphi^a$  define a map $\varphi : {\cal M} \to {\cal E}_{N}$ and the scalar fields kinetic term in the action below is just the pull-back  of the line element of the target space. The action of the 
	MSEGB gravity is then given by 
	
	\begin{eqnarray}
	S=&&\frac{1}{16\pi G}\int d^4x \sqrt{-g} 
	\Big[R -  2g^{\mu\nu}\gamma_{ab}(\varphi)\nabla_{\mu}\varphi^{a}\nabla_{\nu}\varphi^{b} - V(\varphi) 
	+ \lambda^2 f(\varphi){\cal R}^2_{GB} \Big] ,\label{eq:quadratic}
	\end{eqnarray}
	where $R$ is the Ricci scalar with respect to the spacetime metric $g_{\mu\nu}$,   $V(\varphi)$ is the potential of the scalar fields $\varphi=(\varphi^1,...,\varphi^N)$, the coupling function  $f(\varphi)$ depends only on $\varphi$, $\lambda$ is the Gauss-Bonnet coupling constant having  dimension of $length$ and ${\cal R}^2_{GB}$ is the Gauss-Bonnet invariant\footnote{The Gauss-Bonnet invariant is defined by ${\cal R}^2_{GB}=R^2 - 4 R_{\mu\nu} R^{\mu\nu} + R_{\mu\nu\alpha\beta}R^{\mu\nu\alpha\beta}$ where $R$ is the Ricci scalar, $R_{\mu\nu}$ is the Ricci tensor and $R_{\mu\nu\alpha\beta}$ is the Riemann tensor}.  
	
	In order to study this problem we have to specify the theory, i.e. choose a specific form of the functions ${\cal E}_{N}$, $\gamma_{ab}(\varphi)$, $V(\varphi)$ and $f(\varphi)$. Here, as  mentioned, we shall consider MSEGB gravity whose target space manifold is a 3-dimensional symmetric space, namely 
	$\mathbb{S}^3$, $\mathbb{H}^3$ or $\mathbb{R}^3$
	with the metric 
	\begin{eqnarray}
	\gamma_{ab}(\varphi)d\varphi^a d\varphi^b= a^2\left[d\chi^2 + H^2(\chi)(d\varTheta^2 + \sin^2\varTheta d\Phi^2) \right],
	\end{eqnarray}
	where $a>0$ is a constant  and $\varTheta$ and $\Phi$ are the standard angular coordinates on the 2-dimensional sphere $\mathbb{S}^2$.
	The target space metric function $H(\chi)$ is given by $H(\chi)=\sin\chi$ for a spherical geometry, $H(\chi)=\sinh\chi$ for a hyperbolic geometry and
	$H(\chi)=\chi$ in the case of a flat geometry. The parameter $a$ is related to the curvature $\kappa$ of  $\mathbb{S}^3$ and $\mathbb{H}^3$ and we have $\kappa=1/a^2$ for spherical  and $\kappa=-1/a^2$ for hyperbolic geometry.   Our choice of  the target spaces is motivated by the fact that the round $\mathbb{S}^3$, $\mathbb{H}^3$ or $\mathbb{R}^3$ are among the simplest target spaces admitting spherically symmetric  black hole solutions for the ansatz defined below. In addition we shall consider theories for which the coupling function $f(\varphi)$ and the potential $V(\varphi)$ depend on $\chi$ only. This allows the equations for $\Theta$ and $\Phi$ to separate form the main system and  guarantees that the spacetime metric will be spherically symmetric  for the ansatz defined below. 
	
	Instead of making the simplest choice for which all the scalar fields depend 
	on the radial coordinate $r$ only, we   choose here a nontrivial map $\varphi : {\cal M} \to {\cal E}_N$ defined as follows.
	We  assume that the field $\chi$ depends on the radial coordinate $r$, i.e. $\chi=\chi(r)$, and the fields $\varTheta$ and $\Phi$ are independent from $r$ and are given by $\Theta=\theta$ and $\Phi=\phi$ \cite{Doneva_2020}. Our ansatz is  compatible with the spherical symmetry and one can  check that the equations for  $\varTheta$ and $\Phi$ are satisfied.

	In the present paper we are interested in the static and spherically symmetric black hole solutions to the equations of MSEGB gravity
	with a metric 
	\begin{eqnarray}
	ds^2= - e^{2\Gamma}dt^2 + e^{2\Lambda}dr^2 + r^2(d\theta^2  + \sin^2\theta d\phi^2),
	\end{eqnarray} 
	where $\Gamma$ and $\Lambda$ depend on the radial coordinate $r$ only.

	For simplicity, in what follows we shall consider the case with $V(\varphi)=0$. With this ansatz  for the scalar fields and using the above form of the metric, we obtain the following  reduced field equations. 
	\begin{eqnarray}
	&&\frac{2}{r}\left[1 +  \frac{2}{r} (1-3e^{-2\Lambda})  \Psi_{r}  \right]  \frac{d\Lambda}{dr} + \frac{(e^{2\Lambda}-1)}{r^2} 
	- \frac{4}{r^2}(1-e^{-2\Lambda}) \frac{d\Psi_{r}}{dr} \nonumber \\ 
	&& \hspace{0.5cm} - a^2\left[ \left( \frac{d\chi}{dr}\right)^2 + 2e^{2\Lambda}\frac{H^2(\chi)}{r^2}\right] = 0, \label{DRFE1}\\ && \nonumber \\
	&&\frac{2}{r}\left[1 +  \frac{2}{r} (1-3e^{-2\Lambda})  \Psi_{r}  \right]  \frac{d\Gamma}{dr} - \frac{(e^{2\Lambda}-1)}{r^2} - a^2\left[ \left( \frac{d\chi}{dr}\right)^2 - 2e^{2\Lambda}\frac{H^2(\chi)}{r^2}\right] = 0,\label{DRFE2}\\ && \nonumber \\
	&& \frac{d^2\Gamma}{dr^2} + \left(\frac{d\Gamma}{dr} + \frac{1}{r}\right)\left(\frac{d\Gamma}{dr} - \frac{d\Lambda}{dr}\right)  + \frac{4e^{-2\Lambda}}{r}\left[3\frac{d\Gamma}{dr}\frac{d\Lambda}{dr} - \frac{d^2\Gamma}{dr^2} - \left(\frac{d\Gamma}{dr}\right)^2 \right]\Psi_{r} 
	\nonumber \\ 
	&& \hspace{0.5cm} - \frac{4e^{-2\Lambda}}{r}\frac{d\Gamma}{dr} \frac{d\Psi_r}{dr} + a^2\left(\frac{d\chi}{dr}\right)^2 = 0, \label{DRFE3}\\ && \nonumber \\
	&& \frac{d^2\chi}{dr^2}  + \left(\frac{d\Gamma}{dr} \nonumber - \frac{d\Lambda}{dr} + \frac{2}{r}\right)\frac{d\chi}{dr} - \frac{2\lambda^2}{a^2r^2} \frac{df(\chi)}{d\chi}\left\{(1-e^{-2\Lambda})\left[\frac{d^2\Gamma}{dr^2} + \frac{d\Gamma}{dr} \left(\frac{d\Gamma}{dr} - \frac{d\Lambda}{dr}\right)\right]   \right. \nonumber \\
	&& \left. \hspace{0.5cm}  + 2e^{-2\Lambda}\frac{d\Gamma}{dr} \frac{d\Lambda}{dr}\right\} =  \frac{2}{r^2} H(\chi)\frac{dH(\chi)}{d\chi}e^{2\Lambda} \label{DRFE4}
	\end{eqnarray} 
	with 
	\begin{eqnarray}
	\Psi_{r}=\lambda^2 \frac{df(\chi)}{d\chi} \frac{d\chi}{dr}.
	\end{eqnarray}
	
	In order for the above system of equations to describe a black hole the following boundary and regularity conditions have to be satisfied.
	As usual the asymptotic flatness imposes  
	\begin{eqnarray}
	\Gamma|_{r\rightarrow\infty} \rightarrow 0, \;\;  \Lambda|_{r\rightarrow\infty} \rightarrow 0,\;\; \chi|_{r\rightarrow\infty} \rightarrow 0\;\;.   \label{eq:BH_inf}
	\end{eqnarray} 
	
	The very existence of black hole horizon at $r=r_H$ requires 
	\begin{eqnarray}
	e^{2\Gamma}|_{r\rightarrow r_H} \rightarrow 0, \;\; e^{-2\Lambda}|_{r\rightarrow r_H} \rightarrow 0. \label{eq:BH_rh}
	\end{eqnarray} 
	
	By expanding the field equations in series around the black hole horizon we derive the following quadratic equation for the first derivative of the scalar field on the horizon $(d\chi/dr)_H$
		
		\begin{eqnarray}
		&&\left(4\lambda^2 \left(a^2H(\chi_H)^2 - \frac{1}{2}\right) \left(\frac{df(\chi_H)}{d\chi}\right) r_H^3 + 8H(\chi_H)\left(\frac{dH(\chi_H)}{d\chi}\right) \left(\frac{df(\chi_H)}{d\chi}\right)^2 \lambda^4 r_H \right) \left(\frac{d\chi}{dr}\right)_H^2  \nonumber\\
		&&\quad\nonumber\\
		&&+ \left(\left(2a^2H(\chi_H)^2 - 1\right) r_H^4 + 8H(\chi_H)\left(\frac{dH(\chi_H)}{d\chi}\right) \left(\frac{df(\chi_H)}{d\chi}\right) \lambda^2 r_H^2 \right.\nonumber\\
		&&\quad\nonumber\\
		&&+ \left. 16\lambda^4 a^2 \left(a^2H(\chi_H)^2 - \frac{1}{2}\right) \left(\frac{df}{d\chi}\right)_H^2 H(\chi_H)^2\right) \left(\frac{d\chi}{dr}\right)_H +  2H(\chi_H)\left(\frac{dH(\chi_H)}{d\chi}\right) r_H^3 \nonumber \\
		&&\quad\nonumber\\
		&& - \left(\frac{df(\chi_H)}{d\chi}\right)\lambda^2 \left(\left(2a^2H(\chi_H)^2 - 1\right)^2 - 2\left(2a^2H(\chi_H)^2 - 1\right)\right) = 0  \label{eq:initial_cond_dchidr}
		\end{eqnarray}
		
	This equation has two roots for $(d\chi/dr)_H$ but only the one with a positive sign in front of the discriminant gives the Schwarzschild solutions as a limiting case, and therefore this is the one we are adopting. A real root $(d\chi/dr)_H$  exists if the discriminant is positive, which leads to the following inequality
	\begin{eqnarray}
	&\left( {a^2}{{H^2(\chi_H)}}-\frac{1}{2} \right) ^{2} \left( {a^4}{{H^{4}(\chi_H)}}\left(\frac{df}{d\chi}\right)_{H}^{4}{\lambda}^{8}+\frac{3}{2}{H(\chi_H)}
	{\frac{dH(\chi_H)}{d\chi}}\left(\frac{df}{d\chi}\right)_{H}^{3}{\lambda}^{6} r^2_H+\right.\nonumber\\
	&\quad\nonumber\\
	&\left.\frac{1}{2}\left( {a^2} H^2(\chi_H)-\frac{3}{4} \right) {\lambda}^{4} r^{4}_H
	\left(\frac{df}{d\chi}\right)_{H}^{2}+ \frac{1}{64} r_H^{8} \right) \geq 0. \label{eq:condition_existence_BH}
	\end{eqnarray}
	Therefore, the inequality serves as a condition for the existence of a black hole and practically it turns out that for certain Gauss-Bonnet coupling functions it introduces a minimum radius of the horizon below which no black hole solutions are present. One can easily show that for flat target space geometry it reduces to the equation for the existence of Gauss-Bonnet black holes with a single scalar field (see e.g. \cite{Doneva_2018}).
	
	The functions $\Gamma$ and $\Lambda$ have the usual asymptotics at infinity, namely 
	\begin{equation}
	\Lambda\approx \frac{M}{r} + O(1/r^2), \;\; \Gamma\approx  -\frac{M}{r} + O(1/r^2), 
	\end{equation}
	where  $M$ is the black hole mass. The asymptotic behavior of the scalar field $\chi$ can be obtained from the linearized equation for $\chi$  far away from the black hole and we find   
	\begin{equation}
	\chi\sim \frac{1}{r^2} . 
	\end{equation}
	
	This unusual asymptotic has serious physical consequences. It means that the scalar charge associated with $\chi$ is zero which means that the 
	scalar dipole radiation is strongly suppressed. This is very important given the fact that perhaps one of the strongest constraints on the theories of gravity possessing a scalar degree of freedom come from the indirect observations of gravitational wave emission from neutron stars in compact binaries \cite{Freire:2012mg,Antoniadis:2013pzd,Shao:2017gwu}. In Gauss-Bonnet theories possessing one scalar field it has already been shown that scalarized neutron stars can exist \cite{Doneva:2017duq} and it is natural to expect that this will be also true for the MSEGB gravity under consideration. The scalar field is expected to have the same asymptotic as for the black hole case considered here, which means that the close binary pulsars would not emit scalar dipole radiation and no constraints can be put on the theory on the basis of these observations. This fact would potentially allow for much larger deviations from GR. 
	
	\section{Numerical black hole solutions}
	The numerical solutions are obtained using a shooting method to solve the system of equations (\ref{DRFE1})--(\ref{DRFE4}) with the appropriate boundary conditions at the horizon and infinity, as discussed above. Note that, in addition we have a condition for the existence of black holes, namely eq. \eqref{eq:condition_existence_BH}. The calculations are performed using several different forms of the coupling function $f(\chi)$ which allow the existence of hairy black holes, including scalarized ones, and the three possible forms of the target space metric function $H(\chi)$. 
		
	\subsection{Black holes with scalar hair -- linear and exponential coupling}
	In what follows we impose on the coupling function the condition $f(0)=0$. This can be done because the field equations are invariant under the change $f(\chi) \to f(\chi) + const$. In the subsection we will discuss the results for two coupling functions representing linear coupling  
	\begin{equation}\label{eq:Coupling1}
	f(\chi) = \chi
	\end{equation}
	and  exponential coupling
	\begin{equation}\label{eq:Coupling2}
	f(\chi) = e^{\alpha\chi} -1,
	\end{equation}
	where $\alpha$ is a constant. Such a form of the coupling function was used as well in Einstein-dilaton-Gauss-Bonnet (EdGB) gravity with one scalar field \cite{Pani_2011}. In this case the black hole solutions, when they exist, are always endowed with scalar hair and the zero scalar field (Schwarzschild)  case is not a solution of the field equations, unlike the scalarization discussed in the next section. 
	
	As a matter of fact the two coupling functions are equivalent, up to a constant multiplication factor, in the limit of small scalar field $\chi$. As expected, based on the experience with the EdGB gravity, the qualitative behavior of the solutions is very similar for both cases even for larger $\chi$ and the results differ only quantitatively. That is why in the present section we will present results only for the linear coupling \eqref{eq:Coupling1} and where necessary, comment on the exponential coupling \eqref{eq:Coupling2}. The quantities presented bellow are scaled with respect to the coupling constant $\lambda$ in the appropriate way, which effectively leaves us with one free parameter in the theory in the case of linear coupling, namely $a^2$ (for exponential coupling an additional constant $\alpha$ can be introduced in the$f(\chi)$ function that can not be scaled away). 
	
		\begin{figure}[]
		\centering
		\includegraphics[width=0.45\textwidth]{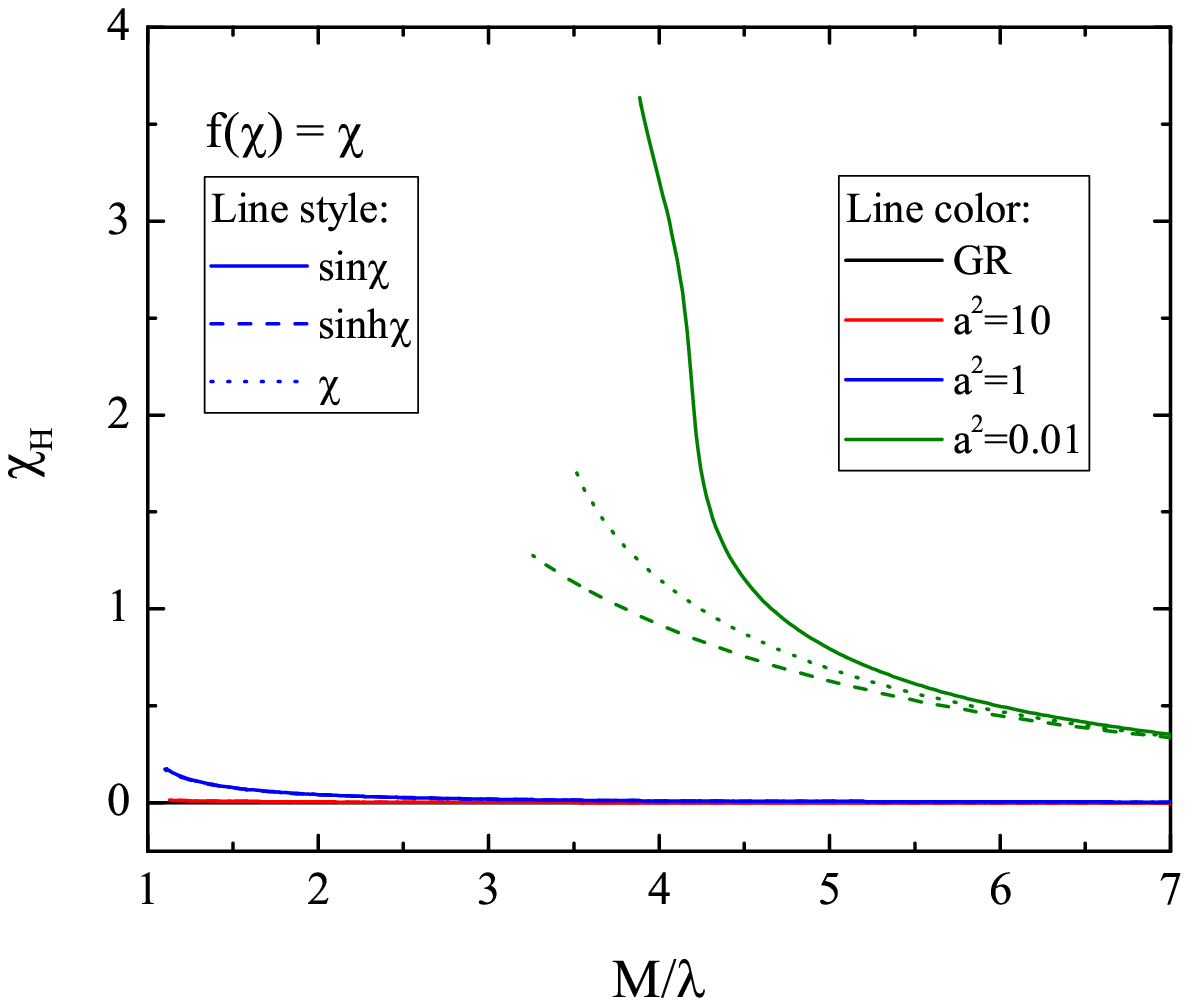}
		\includegraphics[width=0.45\textwidth]{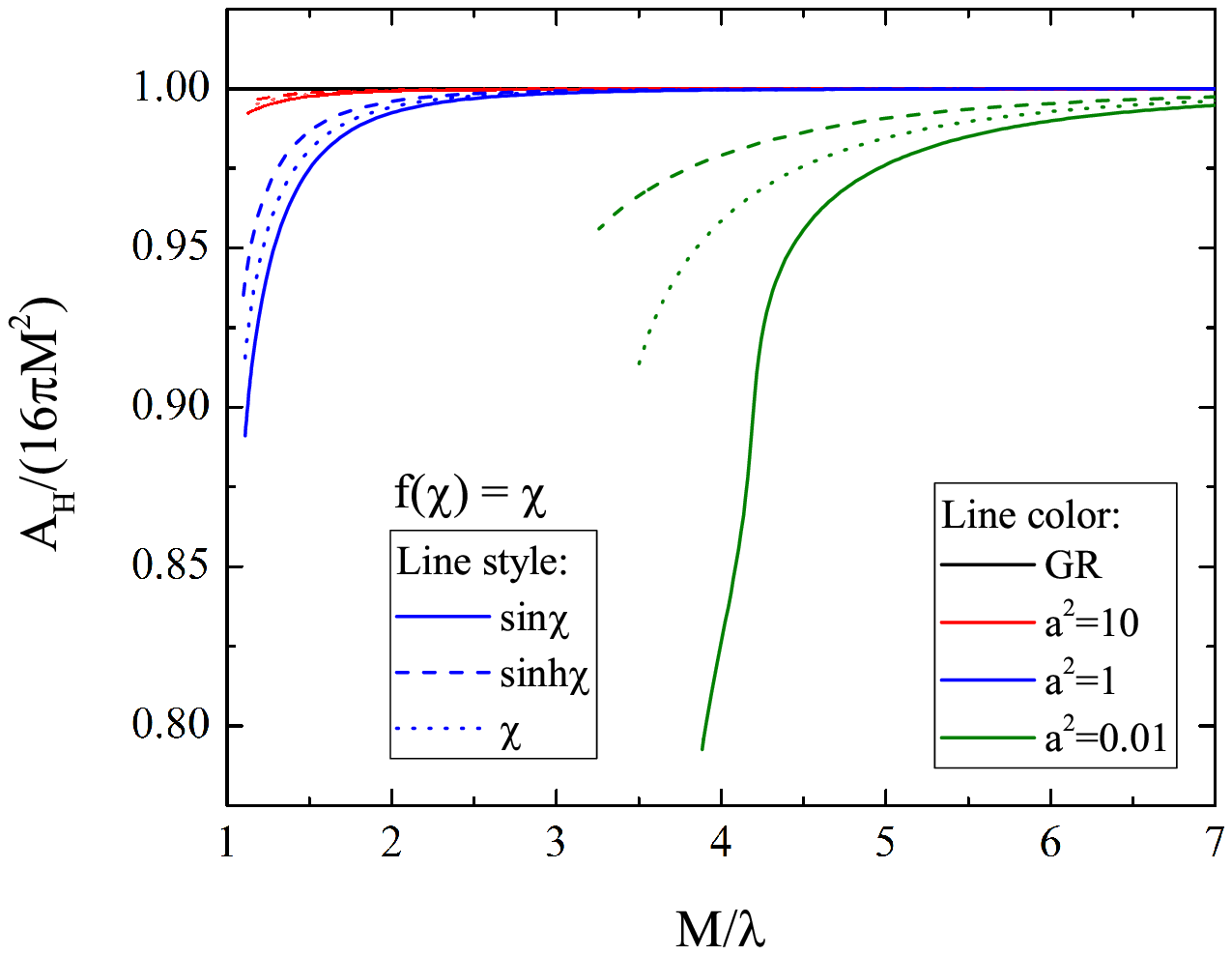}
		\caption{\textit{Left} The value of the scalar field on the horizon as function of the normalized black hole mass.   \textit{Right} The normalized to the Schwarzschild limit area of the black hole horizon $A_H/(16\pi M^2)$ as function ot the mass. The coupling function is $f(\chi)=\chi$ and different colors and styles of the lines corresponds to different choices of $a^2$ and  $H(\chi)$ respectively. The sequences of black holes are terminated at the point where the existence condition \eqref{eq:condition_existence_BH}  is violated. }
		\label{Fig:chi_rh_Ah}
	\end{figure}

	The scalar field on the horizon $\chi_H$ as a function of the normalized (with respect to $\lambda$) black hole mass for the linear coupling \eqref{eq:Coupling1}  is plotted in the left panel of Fig. \ref{Fig:chi_rh_Ah} for different combinations of the parameter $a^2$ and different target space metric functions $H(\chi)$. The scalar field is stronger for smaller masses while, for large $M$ it rapidly tends to zero. Moreover, smaller values of $a^2$ lead to a substantial decrease of  $\chi_H$. For a fixed $a^2$, larger value of $\chi_H$ are achieved for $H(\chi)$ describing spherical geometry, while we have the smallest $\chi_H$ for hyperbolic geometry. Naturally, larger  $\chi_H$ would translate to larger deviations from general relativity and this can be observed in the right panel of Fig.  \ref{Fig:chi_rh_Ah}  where the normalized area of the horizon is plotted as a function of the mass. The normalization of $A_H$ is with respect to the Schwarzschild black hole horizon area and the pure GR case corresponds to the horizontal solid black line at $A_H/(16\pi M^2)=1$. 
	
	The sequences of black hole solutions are terminated at some fixed mass where the condition for the existence of black holes \eqref{eq:condition_existence_BH} is violated and in general, larger $a^2$ lead to a smaller cutoff mass below which no black holes exist. For a fixed branch of solutions, the largest deviation is achieved in the vicinity of this cutoff mass and for the considered range of $a^2$ the difference with the Schwarzschild horizon area is up to $20\%$, but it will increase further for smaller $a^2$. For larger values of $M$ the branches of black holes with nontrivial scalar field practically merge with the Schwarzschild one.  As a matter of fact, we have studied black hole models within a much larger range of $a^2$, namely $a^2 \in [10^{-4},10^2]$, and while the qualitative conclusions remain the same, the quantitative deviations from Schwarzschild increase (decrease) for smaller (larger) $a^2$. Thus, for small enough $a^2$ we can have deviations from GR that are potentially observable while in the large $a^2$ regime the solutions tend to the GR ones and the differences are negligible. Our studies show as well, that the  Gauss-Bonnet gravity with one scalar field and the same coupling function produces deviations that are of the same order as the ones presented in the graphs.

	\begin{figure}[]
		\centering
		\includegraphics[width=0.45\textwidth]{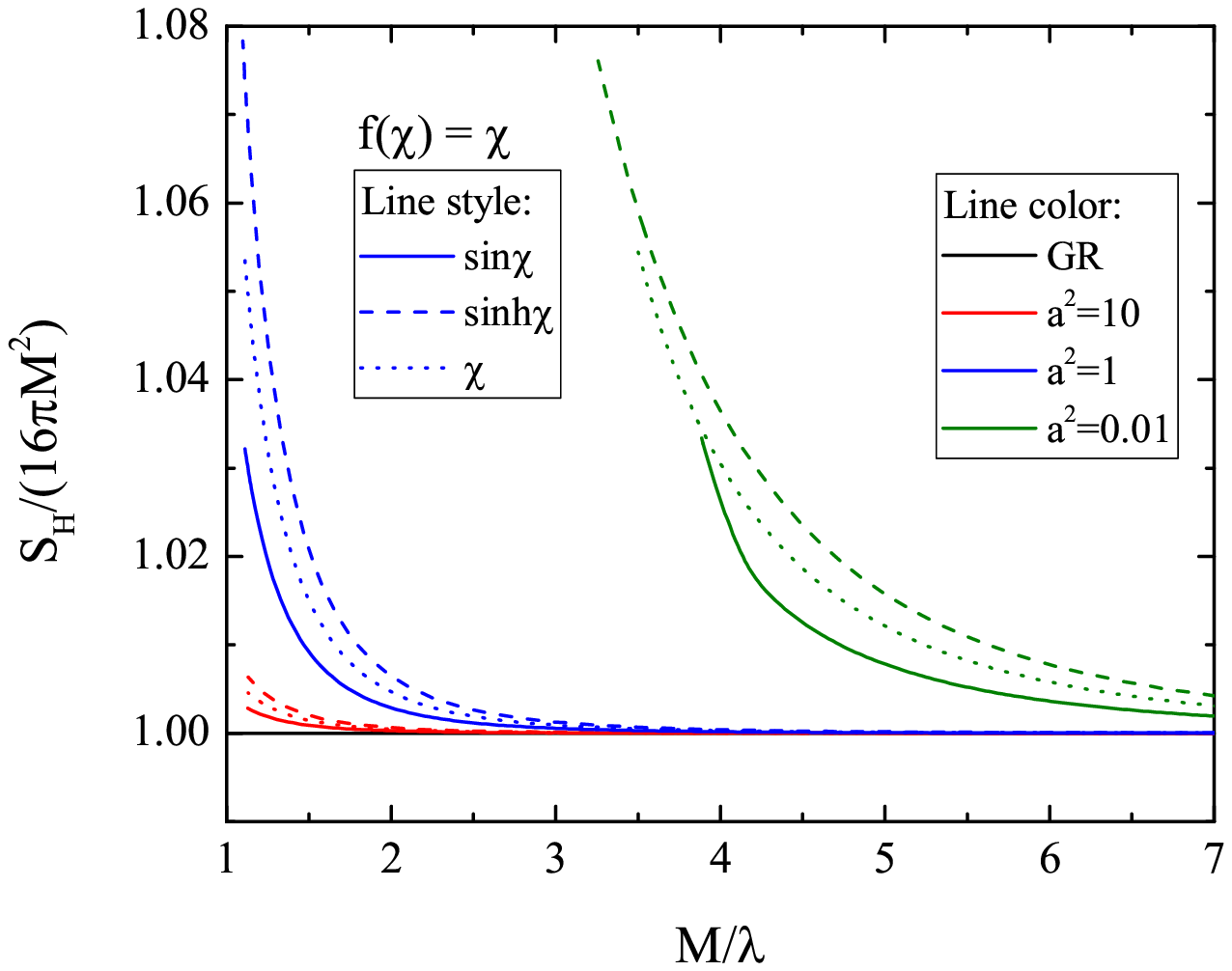}
		\includegraphics[width=0.45\textwidth]{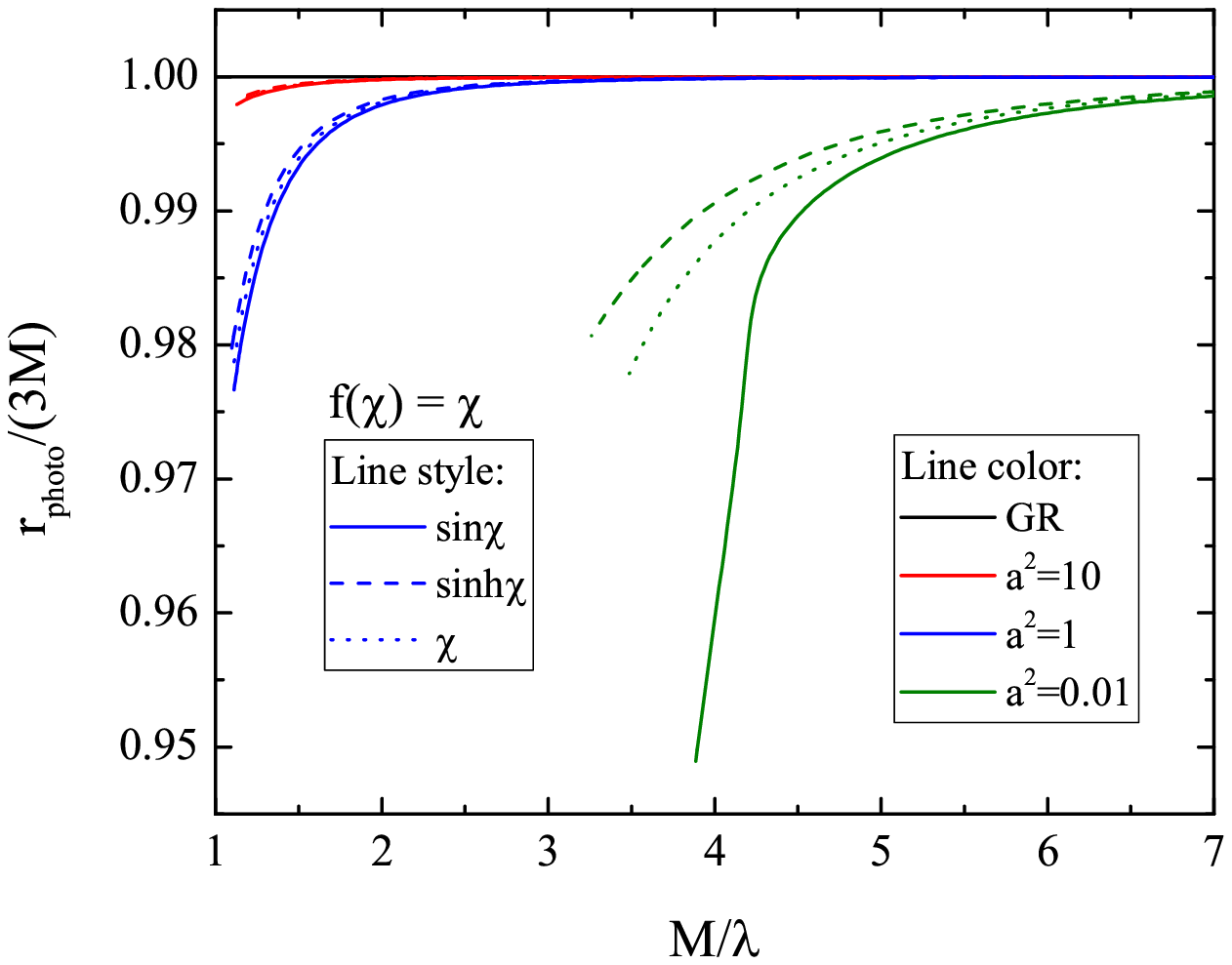}
		\caption{\textit{Left} The normalized, to the Schwarzschild limit, black hole entropy $S_H/(4\pi M^2)$ as function of the mass.   \textit{Right} The normalized, to the Schwarzschild, limit radius of the photon sphere as function of the black hole mass. The coupling function is $f(\chi)=\chi$ and the notations are the same as in Fig. \ref{Fig:chi_rh_Ah}.}
		\label{Fig:chi_Sh_PhS}
	\end{figure}

	The entropy of the black holes can be calculated by using the well-known Wald's formula \cite{Wald_1993}, namely 
	\begin{equation}\label{eq:entropy}
	S_H = \frac{1}{4}A_H + 4\pi\lambda^2 f(\chi_H). 
	\end{equation}
	The entropy, normalized to the Schwarzschild limit $4 \pi M^2$, is plotted in the left panel of Fig. \ref{Fig:chi_Sh_PhS} where the Schwarzschild value is reproduced in the limit of $\chi_H=0$. The black holes with scalar hair always have entropy that is larger compared to the Schwarzschild one. 
	
	Another quantity, that will be discussed, is the radius of the photon sphere $r_{\rm ph}$ defined as the point where the following equality is satisfied
	\begin{equation}\label{eq:photon_sphere}
	\left.\frac{d\Phi}{dr}\right|_{r=r_{\rm ph}} - \frac{1}{r_{\rm ph}}=0.
	\end{equation}
	$r_{\rm ph}$ is directly connected to many observational properties of black holes, such as the frequencies of the quasinormal mode ringing, the black hole shadow and the strong lensing. The radius of the photon sphere is plotted in the right panel of Fig. \ref{Fig:chi_Sh_PhS}, where  $r_{\rm ph}$  is normalized to the radius of the Schwarzschild photon sphere. As we can see, even though for the presented solutions the area of the horizon can differ significantly from the GR case, the deviations in the radius of the photon sphere are quite moderate, up to roughly 5\%. As we have commented, though, smaller values of $a^2$ would lead to larger difference with  the Schwarzschild solution and thus can produce potentially observable effects.
	
	The second coupling function  we have employed has an exponential form and is  given by eq. \eqref{eq:Coupling2}. The qualitative behavior of the solutions is very similar to the first coupling function \eqref{eq:Coupling1}  and the main differences are quantitative ones. That is why we will comment on this case only briefly. For the exponential coupling \eqref{eq:Coupling2} we can introduce an additional parameter in the exponent, namely $\alpha$, that can not be scaled away. Our results show though, that for $\alpha$ of the order of one, and for the same values of $a^2$, the differences with Schwarzschild are a bit smaller compared to the first coupling function.

	\subsection{Scalarized black hole solutions}
	
	In this section we study black hole solutions in MSGB gravity for coupling functions which allow the existence of the zero scalar field (Schwarzschild) solution for all values of the parameters. The Schwarzschild black hole, though, can become unstable below certain mass and spontaneous scalarization is observed, i.e. new branches  of back holes with nontrivial scalar field bifurcate from the GR ones. In order to have scalarized solutions, the following conditions should be satisfied
	\begin{equation}
	\left.\frac{df}{d\chi}\right|_{\chi=0}=0, \;\;\;\; \left.\frac{d^2f}{d\chi^2}\right|_{\chi=0}>0.
	\end{equation}	
	We will discuss two coupling functions satisfying this conditions. 
	
	\subsubsection{Scalarized black holes -- first coupling function}
	
	The first coupling function we will discuss is the following
	
	\begin{equation}\label{eq:Coupling3}
	f(\chi) = \frac{1}{2\beta}\left(1 - e^{-\beta\chi^2}\right),
	\end{equation}
	where we have fixed $\beta = 0.5$. This is exactly the coupling function used in the first study of scalarized Gauss-Bonnet black holes \cite{Doneva_2018} (with a single scalar field) and its advantage is that it leads to nicely behaving branches of stable scalarized black holes. The value of $\beta$ is chosen is such a way that we can have strong deviations from GR and some branches of scalarized black holes can reach close to the $M=0$ limit.
	
		\begin{figure}[]
		\centering
		\includegraphics[width=0.45\textwidth]{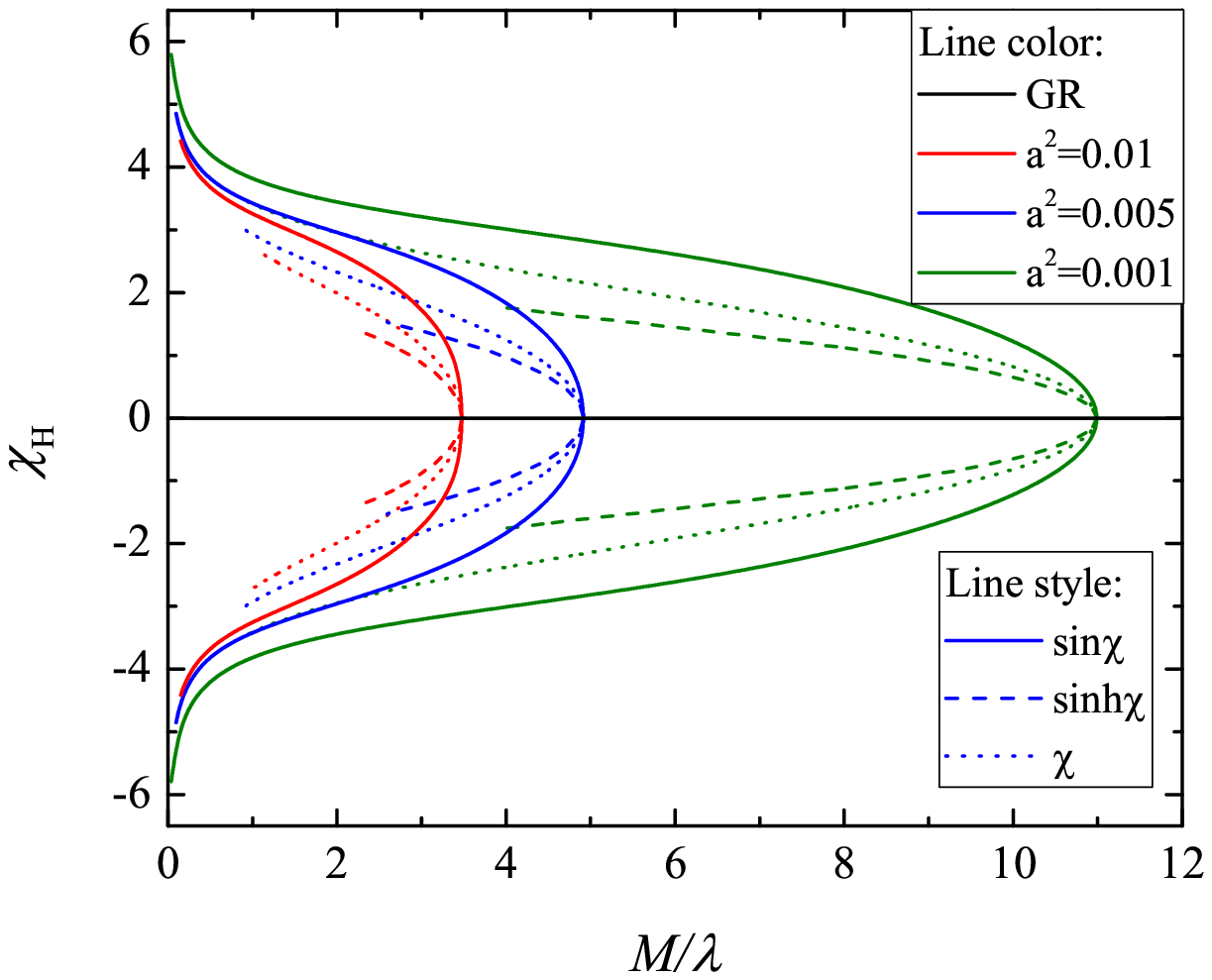}
		\includegraphics[width=0.45\textwidth]{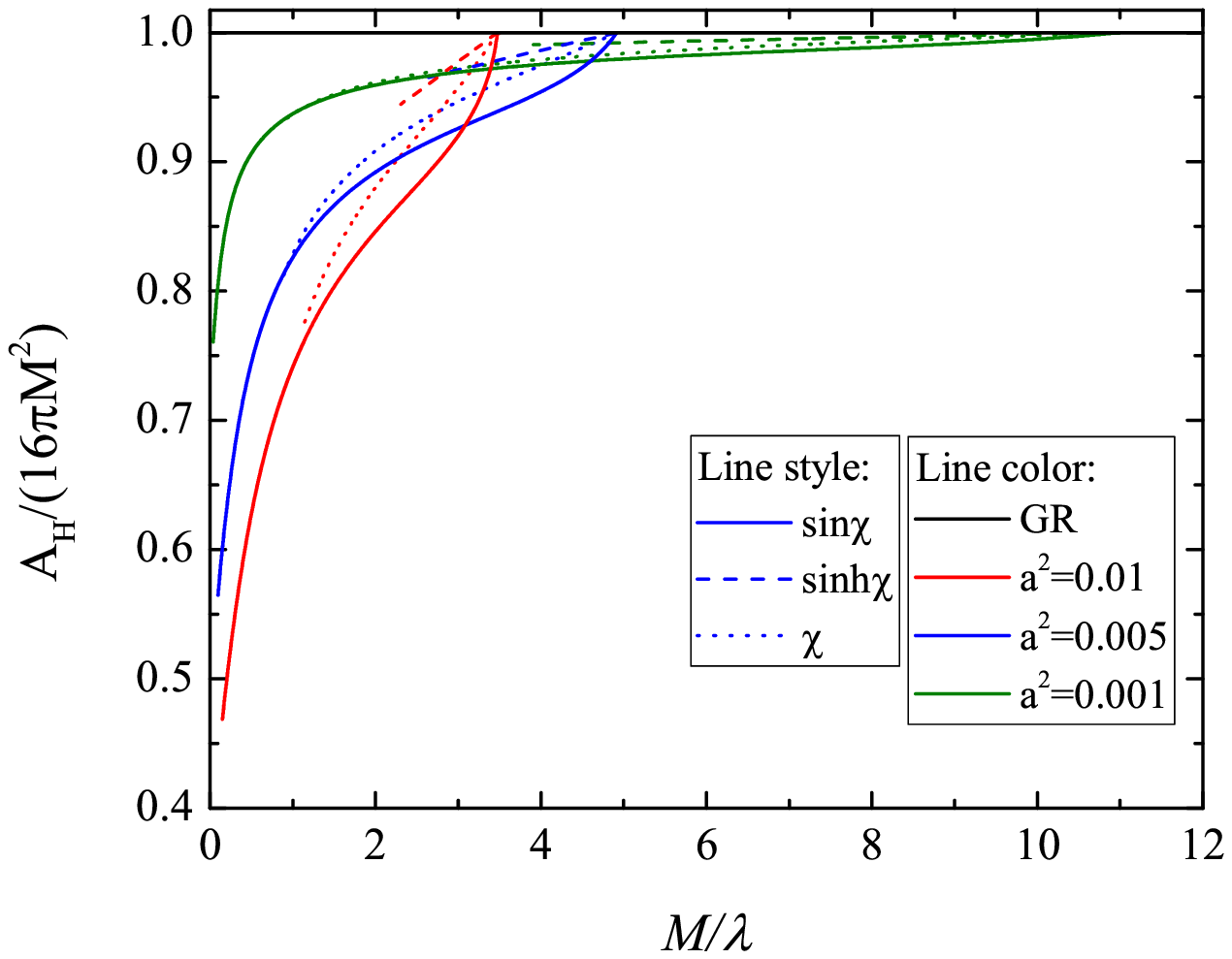}
		\caption{\textit{Left} The value of the scalar field on the horizon as a function of the normalized black hole mass.   \textit{Right} The normalized to the Schwarzschild limit area of the black hole horizon $A_H/(16\pi M^2)$ as a function of the mass. The coupling function is $f(\chi)=\frac{1}{2\beta}\left(1 - e^{-\beta\chi^2}\right)$, where $\beta=0.5$, and different colors and styles of the lines corresponds to different choices of $a^2$ and  $H(\chi)$ respectively.}
		\label{Fig:F1_chih_Ah}
	\end{figure}
	
	In the left panel of Fig. \ref{Fig:F1_chih_Ah} the scalar field at the horizon is shown as a function of the normalized mass for several values of $a^2$ and different forms of the $H(\chi)$ function. The Schwarzschild solutions is depicted with a solid black line at $\chi_H=0$ and it exists for the whole range of parameters. At certain value of the mass, though, the Schwarzschild solution becomes unstable and a new branch of solutions with nontrivial scalar field bifurcates from it. As a matter of fact,  more than one branch of scalarized solutions can exist and these branches can be labeled by the number of zeros of the scalar field. Only the first branch, though, that has no nodes of the scalar field, can be potentially stable \cite{Salcedo_2018,Salcedo_2020} and that is why we will focus only on these solutions. 
	
	Fig. \ref{Fig:F1_chih_Ah} shows that the increase of the parameter $a^2$ shifts the threshold mass where scalarization is observed to lower values of $M$. The deviation of the scalarized black holes with respect to GR can be better judged from the right panel of the figure where the normalized area of the horizon is plotted as a function of the mass. The differences with GR increase with the decrease of the mass. The sequences are terminated either because the condition  \eqref{eq:condition_existence_BH} is violated or because of severe numerical difficulties -- for small masses the equations are becoming increasingly stiff and due to accuracy problem we could not find solutions below a certain small value of $M$. Based on our investigations, though, we believe that this threshold mass is more or less close to the limit where the solutions disappear because of the condition \eqref{eq:condition_existence_BH}. Such numerical problems appear as well for the pure Gauss-Bonnet back holes with a single scalar field \cite{Doneva_2018}. 
	
    The main difference between results for different functions $H(\chi)$ is the threshold masses below which we could not find scalarized solutions. Since larger differences with Schwarzschild occur for smaller $M$, the branches with smaller threshold mass can differ more significantly from GR. Thus for a fixed $a^2$,  $H=\sin \chi$ deviates most from Schwarzschild while $H=\sinh \chi$  has the smallest difference.

	\begin{figure}[]
		\centering
		\includegraphics[width=0.45\textwidth]{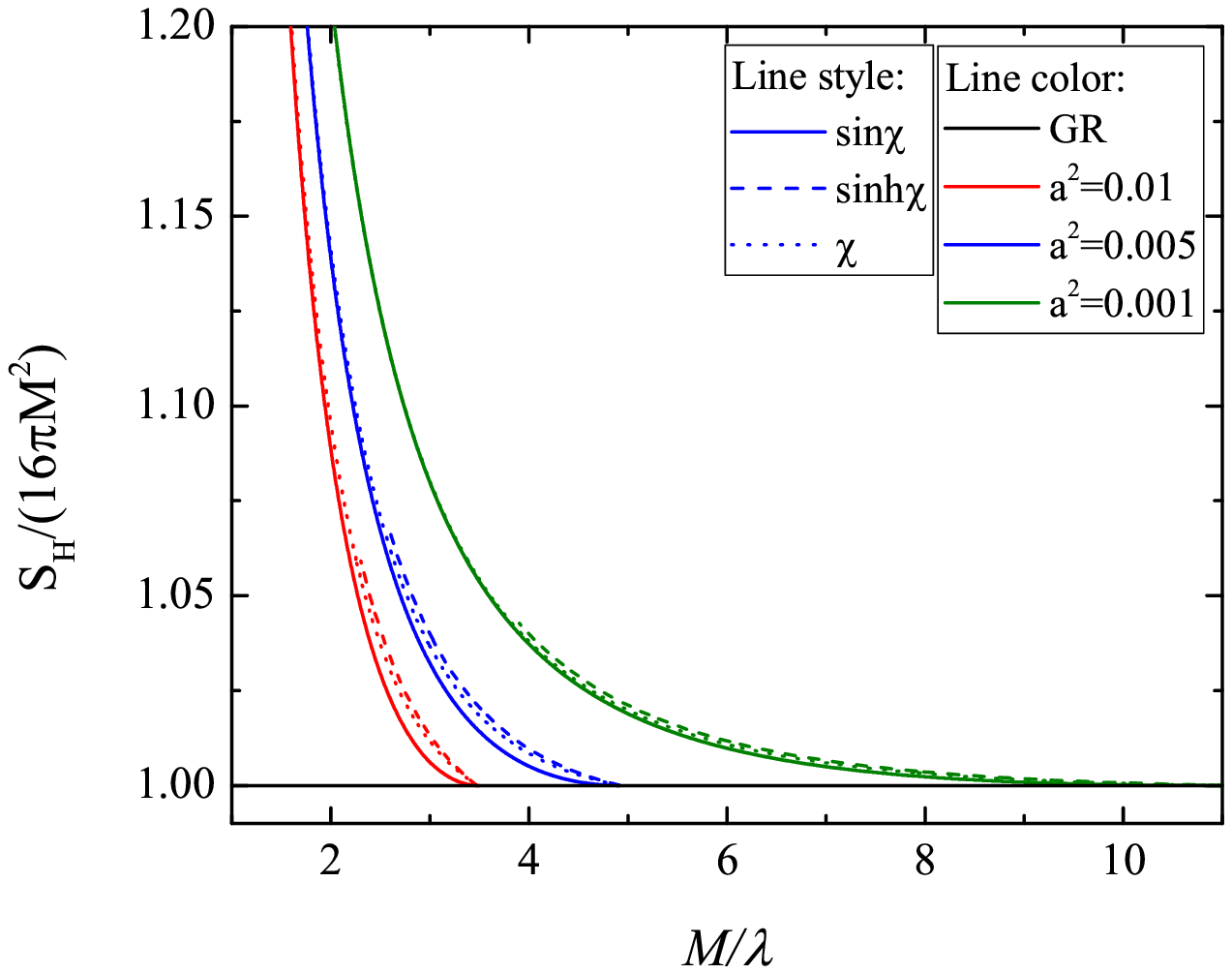}
		\includegraphics[width=0.45\textwidth]{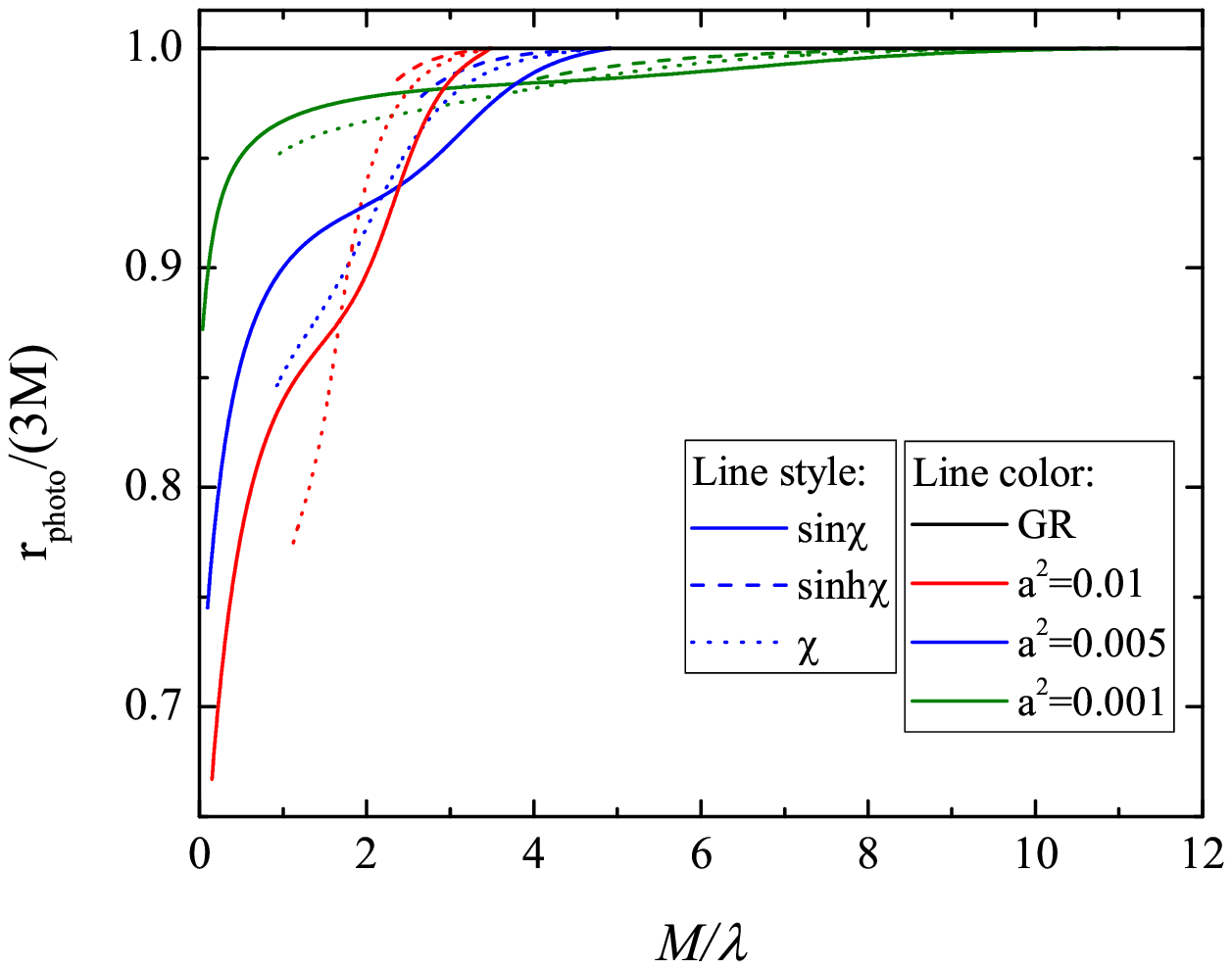}
		\caption{\textit{Left} The normalized, to the Schwarzschild limit, black hole entropy $S_H/(4\pi M^2)$ as function of the mass.   \textit{Right} The normalized, to the Schwarzschild, limit radius of the photon sphere as function of the black hole mass. The coupling function is $f(\chi)=\chi$ and the notations are the same as in Fig. \ref{Fig:F1_chih_Ah}. }
		\label{Fig:F1_SH_Rps}
	\end{figure}

	The next question is whether the new branches of solutions are stable or not. Even without a proper radial perturbation analysis, a good intuition can be obtained from the examination of the entropy -- the solutions with larger entropy are thermodynamically preferred and normally they are the stable ones\footnote{Similar conclusions were made for pure Gauss-Bonnet theories with one scalar field where stable scalarized solutions have also larger entropy than the Schwarzschild one \cite{Doneva_2018,Salcedo_2018}.}. The normalized entropy is plotted in the left panel of Fig. \ref{Fig:F1_SH_Rps}. As one can see the Schwarzschild black hole has always lower entropy than the black holes with scalar hair which gives us the confidence that the scalarized solutions are stable. 
	
	In the right panel of Fig. \ref{Fig:F1_SH_Rps} the radius of the photon sphere is plotted. The differences with GR can be substantial especially for small masses and small values of $a^2$. This can potentially lead to strong imprints of the nonzero scalar field on the astrophysical observations.  
	
	The results up to now were for $\beta=0.5$. We have investigated the solutions for other values of $\beta$ as well and the results remain qualitatively unchanged. The main difference is the quantitative deviation from GR. Loosely speaking, smaller $\beta$ produce larger differences with the Schwarzschild solutions for a fixed black holes mass and fixed $a^2$.

	\subsubsection{Scalarized black holes -- second coupling function}
	
	In this subsection we will focus on a second coupling function that can lead to scalarization:
	\begin{equation}\label{eq:Coupling4}
	f(\chi) = \frac{1}{\beta} \left(e^{\beta \sin^2\chi} -1\right).
	\end{equation}
	The results we will present below are for the case of $\beta=1$ but other choices of $\beta$ are commented as well.
	
\begin{figure}[]
		\centering
		\includegraphics[width=0.45\textwidth]{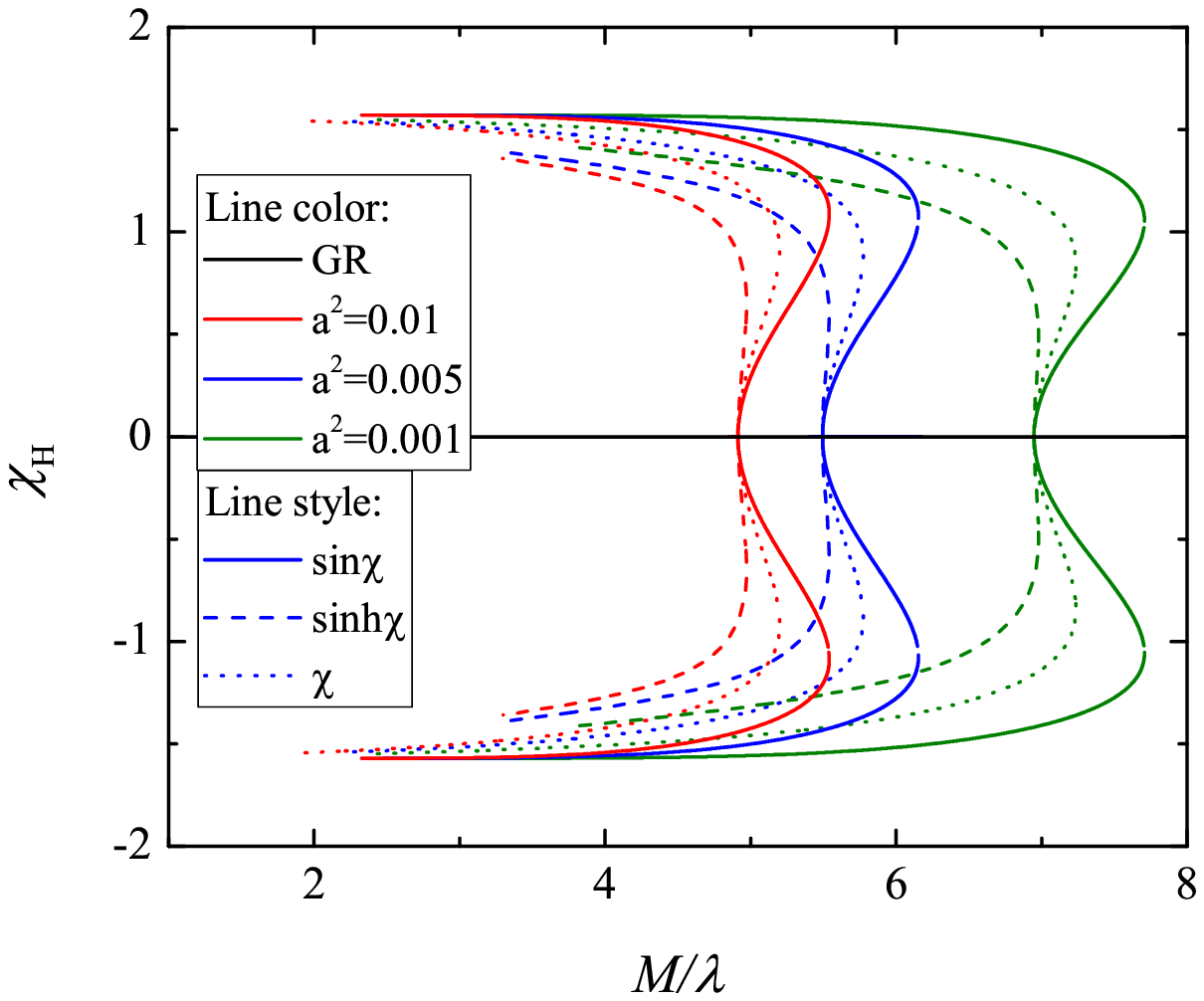}
		\includegraphics[width=0.45\textwidth]{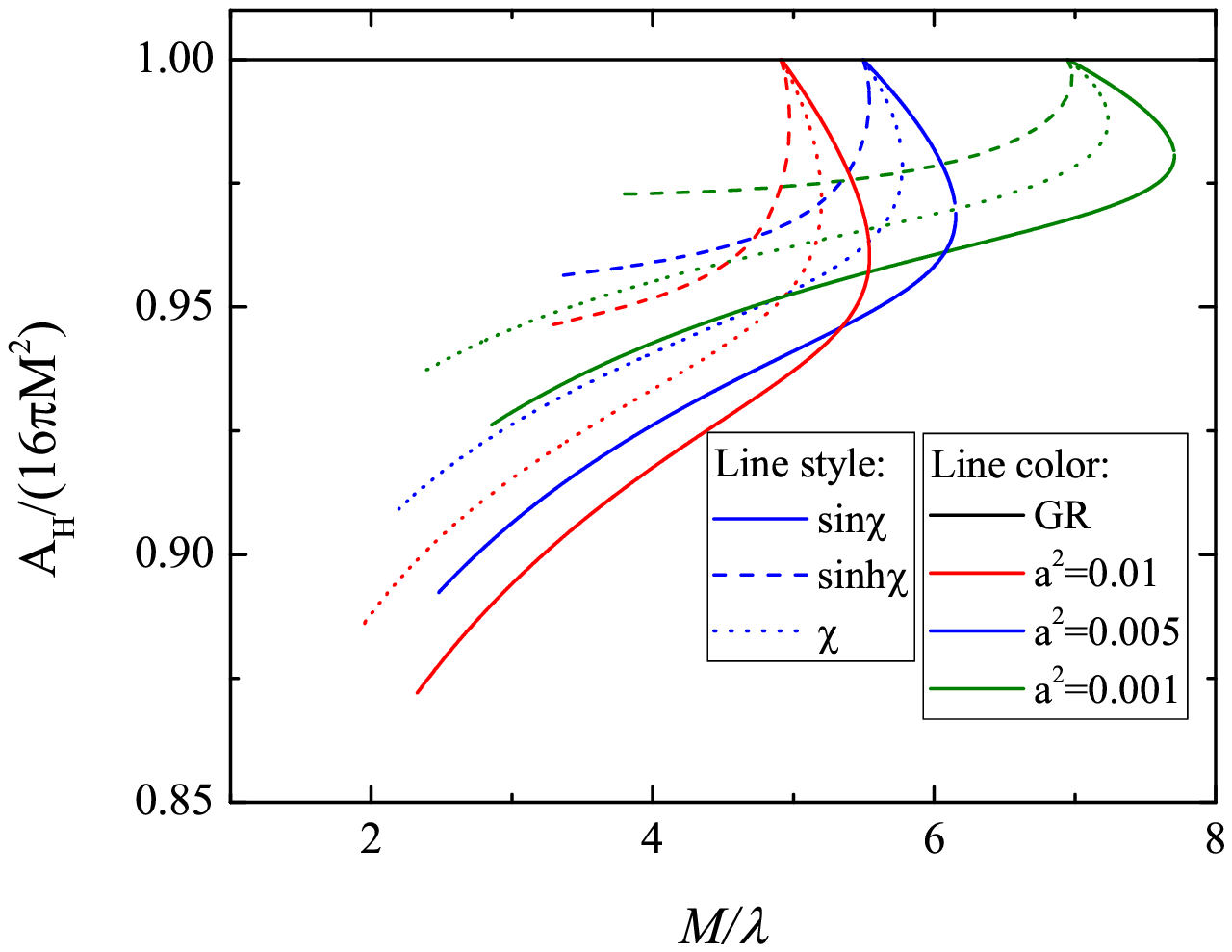}
		\caption{\textit{Left} The value of the scalar field on the horizon as a function of the normalized black hole mass.   \textit{Right} The normalized to the Schwarzschild limit area of the black hole horizon $A_H/(16\pi M^2)$ as a function of the mass. The coupling function is $f(\chi) = (1/\beta) (e^{\beta \sin^2\chi}-1)$, where $\beta=1$, and different colors and styles of the lines corresponds to different choices of $a^2$ and  $H(\chi)$ respectively. }
		\label{Fig:F2_chih_Ah}
	\end{figure}

	The scalar field on the horizon and the normalized area of the horizon are plotted in Fig. \ref{Fig:F2_chih_Ah}. One can observe a very interesting behavior of the scalarized branches -- after the bifurcation, the mass of the hairy black holes first stars to increase and after reaching a maximum it decreases. The branches are terminated at  $\chi_H\rightarrow \pi/2$, since at that point the initial condition for the scalar field \eqref{eq:initial_cond_dchidr} diverges and we could not find any black hole solutions pass this point. Thus, if we choose a black holes mass in the range between the bifurcation point and the maximum of the mass for the scalarized branch, three black hole solutions exist -- two scalarized ones and one zero scalar field Schwarzchild-like solution. Similar features were observed for the first time for scalarized charged black holes with nonlinear electrodynamics \cite{Stefanov_2008,Doneva_2010} and recently also for Gauss-Bonnet black holes with one scalar field when a coupling function with a quartic term is considered \cite{Silva_2019}. While the former case is theoretically very interesting, its astrophysical implications are limited since one requires a nonzero black hole charge. In the latter Gauss-Bonnet case, the difference between the bifurcation mass and the maximum scalarized branch mass was very small for the parameters considered. In addition, the stability of the solutions in this region could not be studied well due to numerical difficulties \cite{Silva_2019}. The results presented here, though, show a clear appearance of a non-negligible region where two scalarized solutions with the same mass coexists and the most important question we have to address is about their stability.

\begin{figure}[]
		\centering
		\includegraphics[width=0.45\textwidth]{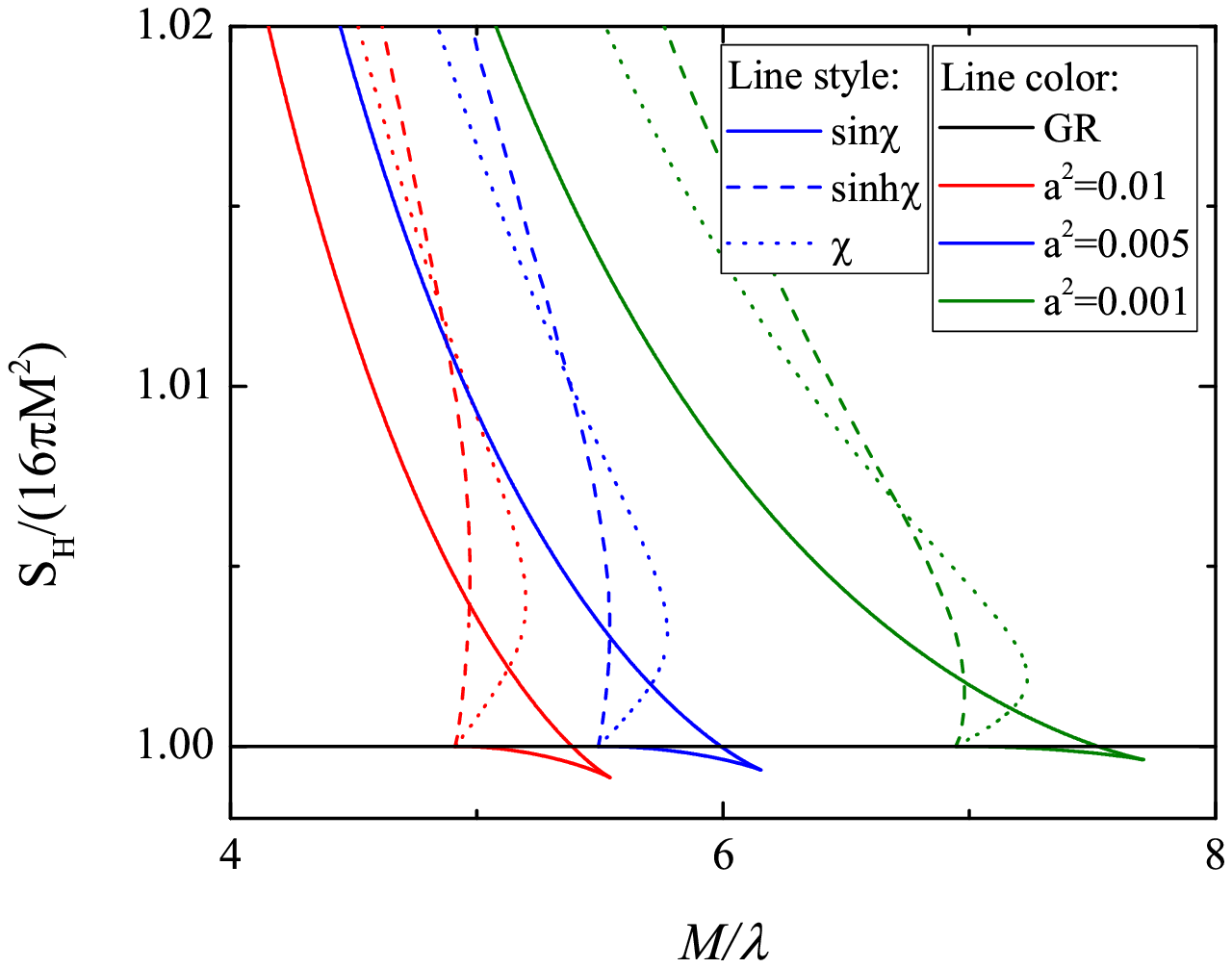}
		\includegraphics[width=0.45\textwidth]{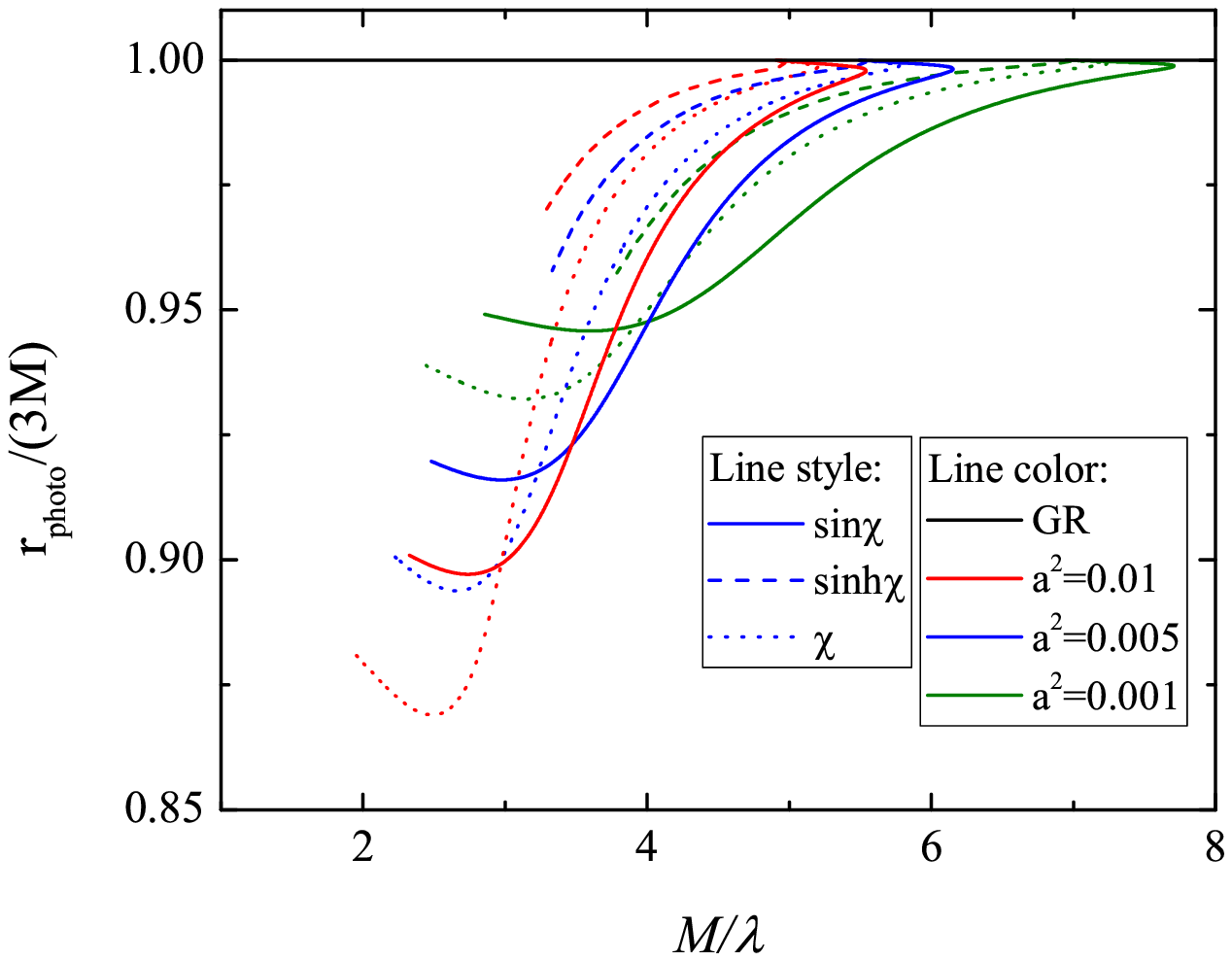}
		\caption{
			\textit{Left} The normalized, to the Schwarzschild limit, black hole entropy $S_H/(4\pi M^2)$ as function of the mass.   \textit{Right} The normalized, to the Schwarzschild, limit radius of the photon sphere as function of the black hole mass. The coupling function is $f(\chi) = (1/\beta) (e^{\beta \sin^2\chi}-1)$, where $\beta=1$ and the notations are the same as in Fig. \ref{Fig:F2_chih_Ah}.}
		\label{Fig:F2_Sh_Rph}
	\end{figure}

	As we commented in the previous subsections, the study of the black hole entropy (\ref{eq:entropy}) can give us strong hints about the (in)stability of the branches. 
	The normalized entropy is depicted in the left panel of Fig. \ref{Fig:F2_Sh_Rph} as a function of mass. Only the region close to the bifurcation point is plotted since this is the most interesting one. For small masses the scalarized branch have entropy larger that the Schwarzschild one and based on the findings in other classes of Gauss-Bonnet theories, we expect that these solutions are stable. With the increase of the mass the scalarized branch reaches a maximum where a cusp on the $S_H(M)$ diagram appears that signals a change of stability. From that point on the scalarized branch is most probably unstable, moreover it has entropy lower than GR. This coincides with what was observed for the charged scalarized black holes in \cite{Stefanov_2008}. An interesting region exists, though, close to the maximum of the mass where the potentially stable part of the scalarized branch has for a small range of masses lower entropy than the Schwarzschild one. Moreover, the Schwarzschild solution is most probably stable there all the way until the bifurcation point. This is a very interesting region and the question about the stability can be rigorously answered only if we perform the linear stability analysis, that will be done in a future publication.

	The normalized radius of the photon sphere is depicted in the right panel of Fig. \ref{Fig:F1_SH_Rps}. The differences with the Schwarzschild one is, as expected, larger for smaller masses and it reach up to roughly 10\%. This would depend of course on the choice of the parameters $\beta$ and $a^2$. If we assume that the middle part of the branch (between the bifurcation point and the maximum of the mass) is indeed unstable, clearly there will be a jump between the last stable Schwarzschild model and the stable scalarized black hole with the same mass. This can potentially lead to interesting observational signatures in scenarios involving dynamical process of scalarization such as the inspiral of compact objects \cite{Palenzuela:2013hsa,Shibata:2013pra,Khalil:2019wyy}.
	
	It is interesting whether such peculiar behavior can be observed for the pure Gauss-Bonnet gravity with one scalar field. Our results show that for the coupling function \eqref{eq:Coupling4} and properly chosen values of $\beta$,  the mass of the scalarized black holes starts to increase after the bifurcation point but the corresponding condition for the existence of scalarized black holes \eqref{eq:condition_existence_BH} is quickly violated and the branches are terminated before a clear maximum of the mass is reached. Of course, careful adjustment of the parameter $\beta$ and/or the coupling function might produce the desired effect, moreover such behavior (but not so well pronounced) was already observed for a coupling function with a quartic term in the scalar field \cite{Silva_2019} .

	At the end of the section let us comment the dependence of the results on the parameter $\beta$ in the coupling function \eqref{eq:Coupling4}. It turns out that the interesting behavior we observed above disappears for small enough $\beta$, while for large $\beta$ the range of masses between the bifurcation point and the maximum of the mass for the scalarized solutions increases. In the case when the non-uniqueness of the scalarized branch disappears the behavior of the solutions is qualitatively the same as for the coupling function \eqref{eq:Coupling3} and that is why we will not comment it further.

	\section{Conclusion}
	
	In the present paper we considered multi-scalar extension of Gauss-Bonnet gravity and focused on models whose target space ${\cal E}_N$ is a 3-dimensional maximally symmetric space, i.e. either $\mathbb{S}^3$, $\mathbb{H}^3$ or $\mathbb{R}^3$. 
	We restrict ourselves to the static and spherically symmetric case with the map $\varphi : {\cal M} \to {\cal E}_3$ explicitly given by
	$\varphi=(\chi(r),\Theta=\theta,\Phi=\phi)$ which is compatible with spherical symmetry. Assuming as well that the 
	coupling function depends only on $\chi$ we proved numerically the existence of black holes in multi-scalar-Einstein-Gauss-Bonnet (MSEGB) gravity. An important property of these solutions is that the scalar field drops at infinity as $1/r^2$ which leads to zero scalar charge and negligible scalar dipole radiation. Thus one can not impose strong observational constraints on the parameters based on the indirect observation of gravitational waves emitter by pulsars in close binaries systems.
	
	We concentrated on several different coupling functions leading to hairy black holes, both scalarized and non-scalarized ones. Since it is possible to rescale with respect to the Gauss-Bonnet coupling parameter $\lambda$ and for a fixed form of the coupling function, we are left with one free parameter $a$ which in the cases of $\mathbb{S}^3$ and $\mathbb{H}^3$ is the curvature radius of the target space. For all examined cases of hairy but non-scalarized black holes we found that in the limit $a^2 \rightarrow \infty$ the results in MSEGB theory converge to the GR ones, and the highest deviations are observed for small values of $a^2$. However, there exist a cutoff mass below which no black hole solutions exist that depends of the parameter $a^2$ and it increases with the decrease of $a^2$. Regarding the dependence of the results on the function $H(\chi)$, the highest deviations from GR are observed for the spherical geometry  and lowest for the hyperbolic geometry. 
	
	The scalarized black holes in MSEGB theory possess the standard features -- at certain mass new branches of solutions with nontrivial scalar field bifurcate from the Schwarzschild one. The point of bifurcation moves towards smaller masses with the increase of the $a^2$. The deviations from GR are strongly dependent on the choice of parameters and larger deviations are observed for smaller $a^2$. The branches either reach zero mass, which results in strong increase of the scalar field and large differences with respect to Schwarzschild for small $M$, or are terminated at some finite mass due to violation of the condition for existence of black holes. For one of the coupling functions we considered, a very interesting phenomena was observed -- the scalarized branch moves first to larger masses and after reaching a maximum $M$, the mass starts to decrease. Thus two subbranches can be distinguish -- a middle one between the point of bifurcation and the maximum mass and an outer branch after the maximum of the mass. Based on thermodynamical studies, we can conclude that most probably the middle branch is unstable, while the outer branch might be stable.
	
	If we assume that the approximate thermodynamical stability analysis coincides with the yet unexplored linear stability of the solutions, then interesting consequences will follow especially for phenomena involving dynamical scalarization of the black holes. The reason is that we would not have a smooth transition between scalarized and non-scalarized solutions, like the standard case where the scalarized branches are potentially stable right from the point of bifurcation where the scalar field tends to zero. In contrast, we will have a jump between the last stable Schwarzschild black hole and the stable scalarized black hole branch. Moreover, there might be a region in the parameter space where stable Schwarzschild black holes coexist with stable scalarized black holes. As a results, one might be able to observe a jump between the scalarized and non-scalarized solutions. For example, this can manifest itself as a sudden change in the gravitational wave frequencies during inspiral.

	In addition we studied the space-time around the obtained black-hole solutions and more precisely  the radius of the photon sphere that is directly related to various astrophysical manifestations of black holes.   In all cases we found that the deviations from GR are more substantial for smaller black hole masses. For the studied forms of the coupling functions and values of the parameters, the largest differences with the Schwarzschild black hole are reached for the scalarized solutions with a maximum deviation of roughly $30\%$. This value would potentially increase if we consider even smaller $a^2$.

	\section*{Acknowledgements}
	DD acknowledges financial support via an Emmy Noether Research Group funded by the German Research Foundation (DFG) under grant
	no. DO 1771/1-1. DD is indebted to the Baden-Wurttemberg Stiftung for the financial support of this research project by the Eliteprogramme for Postdocs. SY would like to thank the University of T\"ubingen for the financial support. KS and SY acknowledge financial support by the Bulgarian NSF Grand KP-06-H28/7. Networking support by the COST actions CA16104 and CA16214 is gratefully acknowledged.
	

\end{document}